# Microscale solute flow probed with rotating microbead trapped in optical vortex

Weronika Lamperska[1,*], Jan Masajada[1], Sławomir Drobczyński[1]

[1]Department of Optics and Photonics, Faculty of Fundamental Problems of Technology,
Wroclaw University of Science and Technology, Wybrzeże Wyspiańskiego 27, 50-370, Wrocław, Poland

*Corresponding author: weronika.lamperska@pwr.edu.pl

**Abstract**

The dynamics of solute flow in the microscopic chamber can be studied with optical tweezers. A method based on the metallic microbeads trapped in the focused optical vortex beam is proposed. This annular beam of a twisted wavefront exerts torque on a reflective object placed inside the dark core of the vortex. The induced rotational movement of the bead is sensitive to local viscosity changes in the surrounding medium, for example during the ongoing dissolution process. Two experimental configurations are described, both relying on tracing the angular velocity of the bead in time. In one-bead configuration the dynamics of local solute concentration can be studied. In two-bead case the direction and speed of solute flow can be probed with a spatial resolution of single micrometers. We approach the elementary problem of sucrose dissolution and diffusion in water. The surprising impression of the reverse solute flow was observed. Further experimental investigation led to the discovery that this phenomenon originates from the sucrose stream-like diffusion in the mid-depth of the measurement chamber. The rotating microbead method applies for various solid and liquid substances and may become a useful technique for microfluidics research.

1. **Introduction**

Optical vortex is a singular beam of a helical wavefront, which means that the phase of the vortex beam circulates around the optical axis [1, 2]. In the cross-section of the vortex phase distribution (Fig. 1a), the equiphase lines converge to a single point, called a vortex point, in which the phase is undetermined. In terms of intensity, the cross-section of the optical vortex is a bright ring with a dark disk in the center (Fig. 1b).

Vortex beams can be used for optical trapping in optical tweezers [3-5]. Transparent particles of the radius smaller than the radius of the vortex beam (e.g. submicron-sized particles) are typically trapped by the bright ring and move along it [6, 7]. On the other hand, the transparent particles large enough to cover the bright ring of the optical vortex beam are trapped at the beam center [6] and so are opaque particles of any size. An example of an opaque particle is a reflective micrometer-sized metallic bead (MB) used in this study. The MB placed inside the dark disk is surrounded by the bright ring of high light intensity and gets repelled from the ring towards its center due to radiation pressure. Thus, the MB is trapped in the dark disk in *x-y* plane and partially along the optical axis *z*.

The twisted curvature of the wavefront results in nonzero angular momentum carried by the optical vortex [3]. This type of angular momentum is called orbital angular momentum (OAM) in contrast to the spin angular momentum (SAM) arising from circular or elliptical polarization of light. In other words, a characteristic, helical spatial distribution of the optical field is responsible for the presence of OAM in optical vortex. Each photon in the vortex beam carries OAM equal to $\pm m\hbar$, where the sign depends on the handedness of the wavefront helix, $m$ is a parameter called topological charge and $\hbar$ is a reduced Planck constant. The higher the vortex charge $m$, the larger the diameter of the bright ring. The transfer of orbital angular momentum between light and matter was observed for the first time by He et al. [8]. After directing a vortex beam onto absorptive CuO particles suspended in water, the particles rotated in the direction determined by the sign of the topological charge of the vortex. Quantitative measurements of the angular momentum transfer were also reported [9-11]. The transfer of OAM can be also achieved with reflective particles such as MBs. The MB illuminated with an optical vortex experiences a constant net torque. Therefore, trapping the reflective particle with a vortex beam not only prevents bead's translation but also causes its rotations (Fig. 1c). The rotational speed depends on beam parameters (laser power, vortex charge), size of the MB and viscosity of the surrounding medium. For fixed beam parameters, the rotating MB becomes a sensitive detector of any changes in viscosity of the medium. In particular, the viscosity growth is indicated by the drop in rotational speed of the trapped bead. The change in viscosity may result from the change in concentration of the solution. The concentration can be changed dynamically, for example during the process of solute diffusion across the sample. We used this approach to study the microscale sucrose diffusion dynamics.

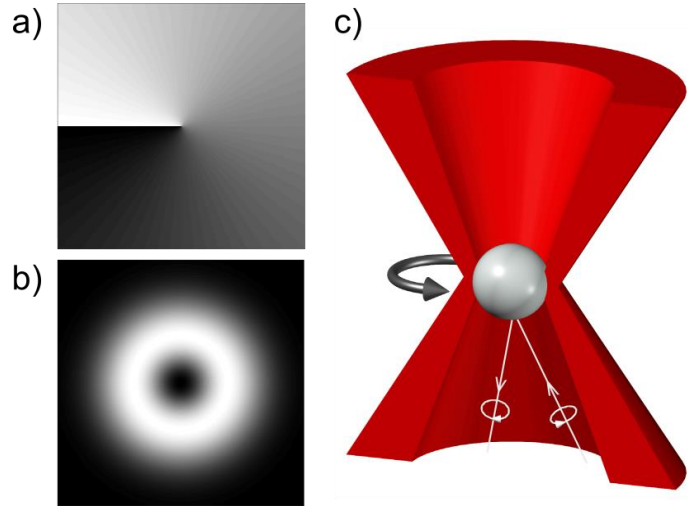

**Fig. 1.** Optical vortex: (a) cross-section of the phase distribution (central point is a vortex point); (b) cross-section of the intensity distribution; (c) schematic view of a metallic bead set into rotations by the focused optical vortex. White arrows represent the incident and reflected rays (for the beam propagating upwards). The rotation direction is reversed upon reflection and thus the nonzero angular momentum is transferred to the trapped bead.

In this paper we report on the measurement method based on vortex-induced rotations of MB and then apply it to study the diffusion dynamics in microscale. We revisit the elementary problem of sucrose dissolution in water in an unconventional manner. Two-laser optical tweezers [12] based on the inverted biological microscope are used to generate vortex traps, manipulate the trapped beads and record the magnified image of the sample. A single rotating MB acts as a microsensor for viscosity changes at the specific point of the sample. Three different phases of MB motion were identified and their background was discussed. The one-MB method proved useful in studying the dynamics of local solute concentration. Furthermore, by using two rotating MBs separated by less than one-tenth of millimeter, one can detect the direction and the propagation speed of sucrose diffusion. The quantity of interest are variations of the MB rotation due to viscosity changes. We report on the nonintuitive observation of distinctive solute flow in the chamber. The origin of this phenomenon was attributed to the sucrose stream-like diffusion in the central depth of the measurement chamber. Two-MB method offers the opportunity to probe the diffusion flow with a spatial resolution of single micrometers. There exist a few methods of studying fluid flows with comparable or greater resolution, such as micro-particle image velocimetry [13, 14], surface plasmon resonance [15, 16], micro-laser Doppler velocimetry [17], positron emission particle tracking [18], NRM [19] and MRI [20]. Most of these techniques require advanced research equipment and are limited to specific experimental conditions (in terms of substrate, substances, physical properties). Here we present a method designed for studying diffusion based on optical tweezers. Nowadays, optical tweezers can be found in many optical laboratories and thus we believe the proposed method could be implemented and tested by a number of research groups.

## 2. Experimental setup

The experiments were performed in two-laser optical tweezers shown in Fig. 2a. In this system there are two optical paths – holographic and non-holographic one. Each path is equipped with a separate laser source emitting the infrared beam of Gaussian intensity distribution. The holographic path is built of Nd:YAG laser (Quantum Ventus, wavelength: 1064nm, 4W) and spatial light modulator (SLM, HoloEye-Pluto). SLM converts the incident Gaussian beam into one or more vortex beams. Vortex charge *m* is determined by the structure of the computer-generated phase hologram displayed on the SLM matrix (Fig. 2b,c). Moreover, combining vortex-generating hologram with the phase diffraction grating (Fig. 2d) enables to control the vortex trap position in the observation plane. The non-holographic path consists of Nd:YAG laser (Quantum Ventus, wavelength: 1064nm, 3W) and a system of two scanning galvanometer mirrors (Thorlabs, GVS002). Electrically addressed mirrors deflect the incident Gaussian beam providing precise beam steering. In both paths half-wave plates are used to set two orthogonal linear polarization states of the beams. The beams are combined by a polarizing beamsplitter (PBS) and then focused by a microscope objective (100x, oil-immersion, NA=1.3, Olympus UPLAN FL N) in the sample plane. The sample plane is illuminated with a halogen lamp, imaged onto the CMOS camera and displayed on the computer screen.

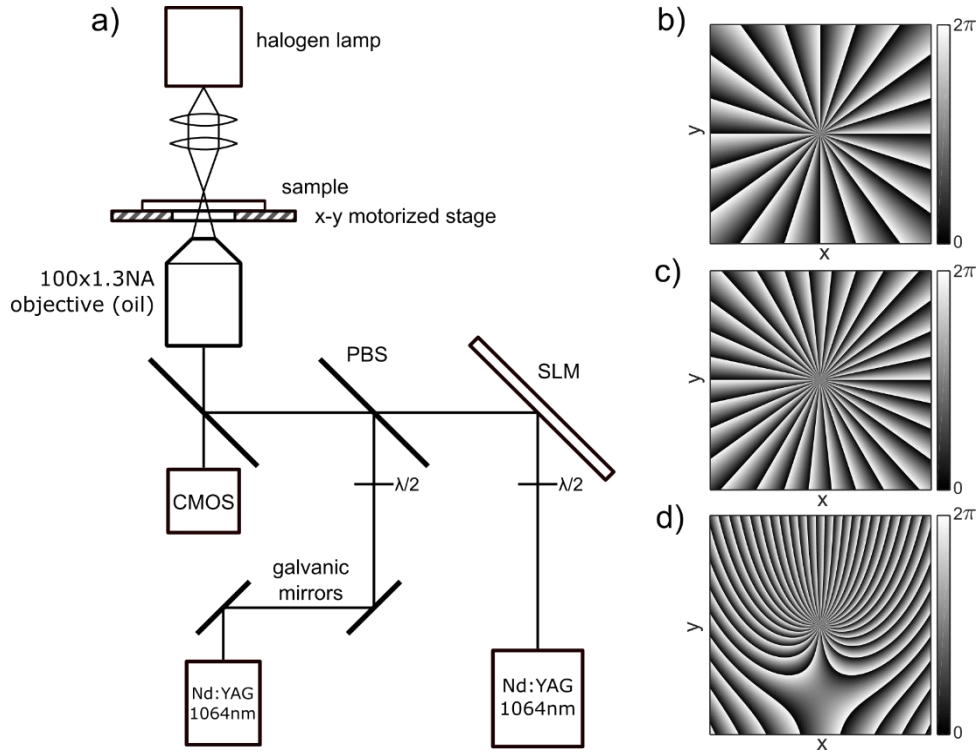

**Fig. 2.** (a) Simplified scheme of two-laser optical tweezers consisting of holographic and non-holographic optical path; λ/2 – half-wave plate, PBS – polarizing beamsplitter, SLM – spatial light modulator; (b-d) exemplary computer-generated phase holograms displayed on the SLM: (b) hologram producing vortex of charge $m=20$; (c) hologram producing vortex of charge $m=30$; (d) the same as in (c) but with added diffraction grating. The position of the optical vortex in the sample plane is controlled by the density and direction of the diffraction grating.

In the sample plane tightly focused beams become optical traps of two kinds – vortex traps from the holographic path and Gaussian traps from the non-holographic path. The holographic beam suffers from substantial power losses at the reflection from the SLM whereas the non-holographic beam leaves the system of galvanic mirrors almost lossless. Thus, despite the higher output power of the vortex-generating laser, vortex traps are weaker than non-holographic traps. Vortex traps are used for trapping the particles in a fixed position and induce their rotations due to angular momentum transfer from the beam to the particle. The non-holographic path supports the experiments with an auxiliary Gaussian trap used for detaching the MBs from the bottom coverslip.

## 3. Materials and methods

The sample consisted of a chamber sandwiched between two coverslips (Fig. 3). The chamber was made by cutting a desired shape in a one-millimeter-high foam tape. Whereas the bottom coverslip covered the entire chamber, the top coverslip left a small gap. At the beginning of each experiment, the chamber was filled with the metallic microbead suspension. During the experiment, the desired substance (either solid or liquid) could be added to the sample through the gap. Chambers of two different geometries were used. The first one was a simple circular chamber (Fig. 3a). According to its name, it had a single circular cavity (10 mm in diameter) of the volume of 78 µl. The second one was a bicircular chamber (Fig. 3b,c) consisting of two circular cavities (both 5 mm in diameter) connected with a channel. The total volume of bicircular chamber was about 41 µl. For the sake of clarity, the cavity with a gap is referred to as cavity A and the fully enclosed cavity is named cavity B.

The metallic microbead suspension was made by suspending 0.5 μl of dry silver-coated soda lime glass microspheres (Cospheric SLGMS-AG, particle size: 5-22 μm) in 1 ml of purified water and adding 0.1 μl of powdered BSA (bovine serum albumin). Metallic beads tend to sink and accumulate at the bottom glass of the sample. BSA was used for reducing the unwanted effect of beads attaching to the glass. Otherwise, it would be challenging to detach the beads with just the use of optical forces. The solution was stored in 4°C and mixed for 5 minutes before use. During the experiments the crystals of sucrose (saccharose, table sugar) were added to the chamber. The average mass of a typical crystal used in experiment was estimated at $(1.0 \pm 0.1)$ mg.

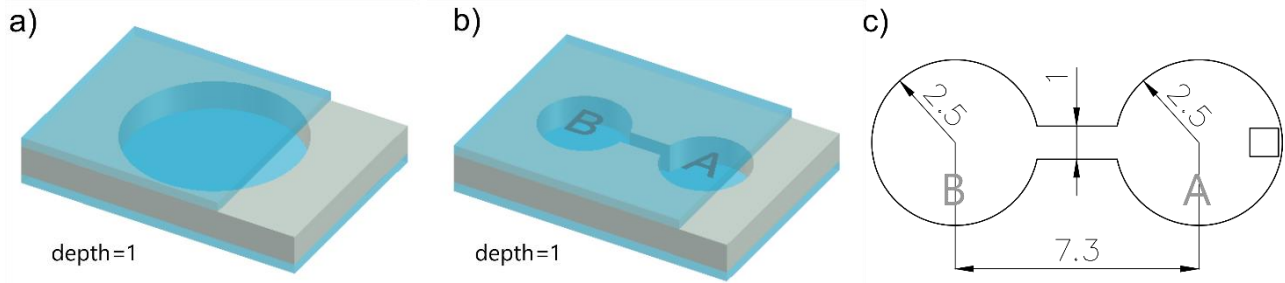

**Fig. 3.** Schematic of the (a) circular and (b) bicircular chambers and (c) dimensions of the bicircular chamber; in (b) and (c) letters A and B denote cavity A and B, respectively. The chambers were cut in the foam tape (gray layer) and placed between two coverslips (blue layers). The top coverslip did not cover the chamber entirely. The narrow gap allowed for adding solid or liquid substances to the chamber during the experiment. The gap width was about 1.5 mm for both types of chambers. The depth was uniform in the entire chamber. The sample thickness was 1.34 mm (1 mm foam tape + 2·0.17 mm coverslips); In (c) a square in cavity A represents one sucrose crystal (approx. a cube of side length 0.9 mm) added through a gap. Units: millimeters.

### 4. Measurement procedure

The measurement procedure in the bicircular chamber was as follows. In the first step, the sample was prepared – the chamber was cut and filled with metallic beads suspension. Then, in the desired region of the chamber, two beads of a similar size (8-12 μm) were trapped in two separate vortex traps. The distance between traps was about 70-80 μm. This separation (7-10 times as large as the bead) is large enough to prevent one bead affecting the other. The vortex charge $m$ was adjusted to the bead size so as to achieve the stable rotational speed (Fig. 4b). For a narrow vortex ring (low charge), the rotations are very fast and may cause escaping the bead from the trap (Fig. 4a). On the other hand, wide ring (high charge) causes slow, nonuniform rotations (Fig. 4c). Since the measurements are based on tracking the change in rotational speed, its absolute value is irrelevant and may be different for both beads. Apart from inducing rotations, optical vortex also serves to lift the bead a few micrometers away from the bottom glass in order to eliminate bead-surface interaction. For the lifting to occur, the radiation pressure imparted on the bead needs to overcome the mass of the bead. In the oversized vortex (Fig. 4c) there is too little radiation pressure inside wide central dark disk and the bead stays at the bottom of the chamber. The vortex charge used for 8-12 μm beads ranged from $m=24$ to $m=30$.

In the third step, the recording of beads rotations starts. As a rule of thumb, after 4-5 minutes of undisturbed motion the rotations were considered stable which was confirmed by the visual inspection. Next, one sucrose crystal is added through the gap at the top of the chamber. As the sucrose dissolves, the viscosity inside the chamber increases. In reaction to the viscous drag growth, the rotations of the beads slow down. Comparing the reaction time for both beads one can find the direction and the speed of sucrose diffusion in the chamber. Fortunately, the beads are not perfectly smooth, but there are some impurities attached to their surface which make the rotations visible (e.g. see Online Resource 1). Therefore, the exact values of the angular velocity can be reconstructed straight from the recording. The movie was processed by hand as the rotational velocities were relatively small (of the order of one rotation per second). In case of faster rotations one would probably need to process the movie automatically with some dedicated software. Depending on the rotational velocity, a proper frame rate of the recording should be provided in order to follow the bead motion. In our experiments the sufficient recording speed was 60 fps.

Although the presented procedure refers to two-bead experiments, a similar treatment is used for the one-bead case (in circular chamber). Instead of two optical vortex traps, only one trap is generated. Since the laser power is no longer divided between two traps, there is greater radiation pressure (lifting force) in the single trap. This allows for using larger beads (12-16 μm) in one-bead experiments. Vortex charge ranged from *m*=30 to *m*=35.

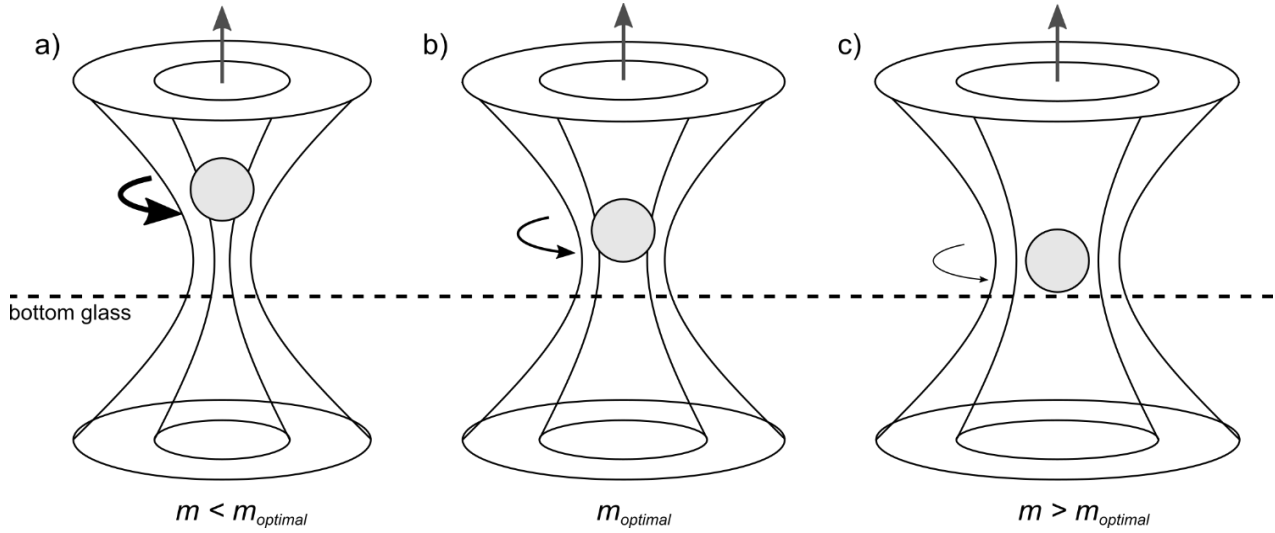

**Fig. 4.** The bead trapped in optical vortex of charge *m*. Vortex ring diameter, rotational speed of the trapped bead and the bead height inside the vortex depend on the vortex charge. For a given bead size there is an optimal charge $m_{optimal}$; (a) $m < m_{optimal}$, narrow vortex ring, rapid bead rotations (thick arrow) and strong lifting; (b) $m = m_{optimal}$, proper vortex width, fast yet well-balanced rotations, bead lifted to height not bigger than its diameter; (c) $m > m_{optimal}$, wide vortex ring, very slow and irregular rotations (thin arrow), no lifting.

## 5. Results

### 5.1. Circular chamber

A single rotating bead was located in the central region of the cavity (Fig. 5a, inset). Subsequent crystals of sucrose were added in 5-minute intervals. There is a characteristic pattern in the graph shown in Fig. 5a. After each addition the rotations of the bead speed up for about 20-40 seconds, then slow down substantially and speed up again until another addition occurs. These three identified phases of bead motion originate from three different effects. The crystal of sucrose put into the chamber triggers a mechanical disturbance in liquid resembling a stone thrown into the water. However, if instead the insoluble substance is added to the chamber, for example a grain of sand, the generated disturbance is many times weaker. It led us to conclusion that the induced mechanical wave is additionally driven by the initiated dissolution. This assumption was confirmed for other soluble substances – glycerol and $KMnO_4$. The mechanical wave propagates uniformly from the crystal location. Thereby, the wave reaching the bead position is directed towards the rear wall of the chamber. The bead gets displaced from the center of the trap and is pushed into the bright ring of optical vortex. Due to the high intensity of light in the ring, angular momentum transfer to the bead increases. Consequently, greater torque is exerted on the bead leading to the observed angular acceleration in the first phase. Simultaneously, there is another wave, the "viscosity wave" of the dissolved sugar. Compared to the mechanical wave, the propagation speed of the viscosity wave is many times lower due to the slow process of dissolution (usually about 2-3 minutes for one crystal to dissolve completely). This explains the delay in bead reaction. The sudden drop in rotational speed indicates the arrival of the high solute concentration at the bead location. In the third phase, a continuous increase in angular velocity is observed. Due to the ongoing diffusion, the sucrose extends to the entire volume of the chamber. At the bead position, local sucrose concentration decreases until the solution becomes homogeneous and angular velocity stabilizes. In Fig. 5a. the stabilization point was not reached because of the insufficient time intervals between subsequent crystal additions.

Changes in angular velocity of the bead can be recalculated into changes in viscosity of the medium. Viscous torque (drag torque) acting upon a rotating bead equals [21]

$$\tau_{drag} = -8\pi\mu r^3 \Omega, \qquad (1)$$

where $\mu$ - viscosity of a medium, r – bead radius, $\Omega$ – angular velocity of a bead. In the described experiment (Fig. 5a) velocity changes were slow (maximal angular deceleration of about 0.004 rad/s²) and nearly linear in all three phases of bead motion. Thus, the entire process is quasi-static and the drag force remains constant. According to Eq.(1) for constant drag torque viscosity is inversely proportional to angular velocity – an increase in viscosity causes a proportional decrease in velocity. Relative viscosity of the medium (Fig. 5b) can be obtained from the measured angular velocity Ω(t) (Fig. 5a) in the following way:

$$\frac{\mu(t)}{\mu_0} = \frac{\Omega_0}{\Omega(t)}, \qquad (2)$$

where $\mu_0$ and $\Omega_0$ are respectively viscosity and angular velocity at the reference point. The reference point was the moment just before the first sucrose crystal addition (188. second of the experiment). Earlier fluctuations of the velocity and thus the calculated viscosity are the artifacts and should be neglected. As mentioned before, time intervals (about 5 minutes) between each addition were too short for velocity to reach a stabilization point. However, relative viscosity just before the next addition should be close to the target values. As shown in Fig. 5b, relative viscosity just before second, third and fourth blue dot lied within or close to the expected values range (blue strips). On the other hand, evaporation of water occurred during the measurement. Water evaporated from the gap region of the chamber until the substrate in the open area got dry. Due to the capillary forces, the fluid between two coverslips did not flood the dry substrate. As the evaporation continued, concave meniscus appeared and then penetrated the fluid layer. It was estimated that the meniscus started to form after about 20 minutes of the experiment. Knowing the evaporation time of the water volume in the open area (about 7.3 μl) one can calculate the modified expected viscosity values of the 1, 2 and 3-crystal solution (red dashed line). For details see Supplementary Information.

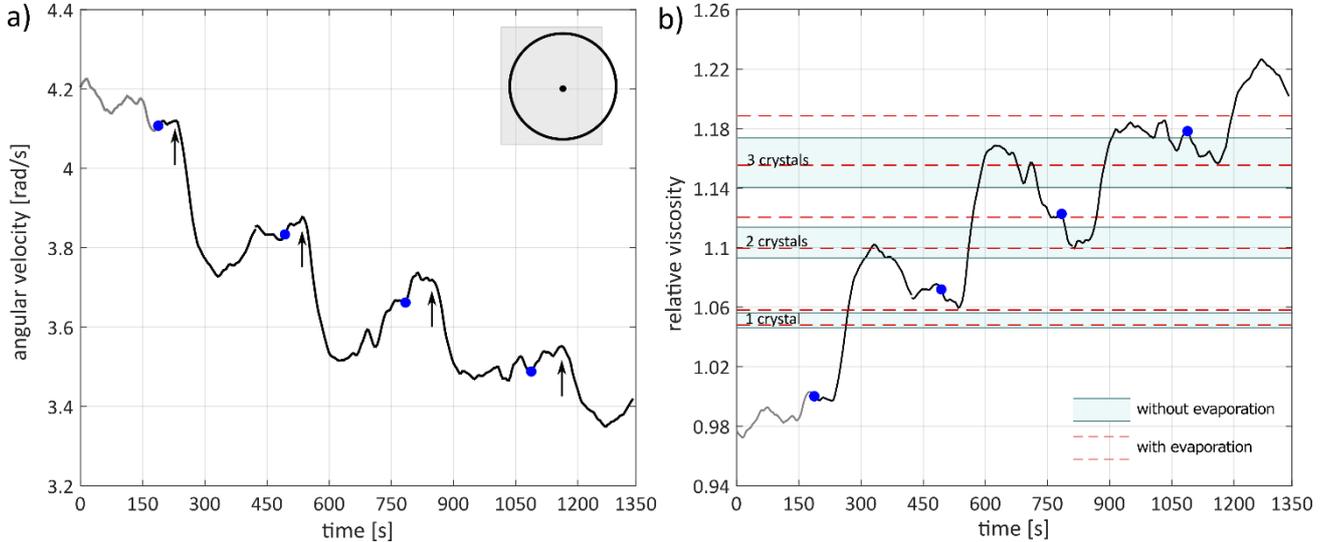

**Fig. 5.** a) Rotational speed of one bead located in the center of the circular chamber. The approximate position of the bead (not to scale) is shown in the graphical legend. The moments of one sucrose crystal additions are marked with blue dots and the moments of bead reaction to the viscosity wave (the beginning of each major drop in velocity) are marked with arrows; b) angular velocity from a) recalculated into relative viscosity change due to the change in concentration. The reference point (for which relative viscosity equals 1) was the moment just before the first sucrose crystal addition (blue dot at 188. second). Earlier fluctuations of the velocity and thus the calculated viscosity (gray line) are the artifacts and should be neglected. Three blue strips represent the expected viscosity values after addition of subsequent sucrose crystals. Mass of each crystal was estimated at $(1.0 \pm 0.1)$ mg. The lower edge of each strip corresponds to light crystals (0.9 mg each) and the upper edge to heavy crystals (1.1 mg each). Blue strips were calculated with neglected evaporation. With evaporation included, the expected viscosities grow and the strips are displaced (edges marked with dashed red line). Details of the calculations can be found in the Supplementary Information.

The major technical difficulty connected with the circular chamber is the strong mechanical disturbance after the addition of sucrose crystal. If the bead gets pushed out of the vortex trap (outside the bright ring), the experiment is disrupted and needs to be repeated. This problem can be resolved with the use of the bicircular chamber, for which the mechanical wave generated in cavity A is separated from the cavity B with the relatively narrow channel. This way the stability of the bead is improved and allows for two-bead experiments.

### 5.2. Bicircular chamber

Initially, the aim of the two-bead experiments in the bicircular chamber was to estimate the propagation speed of the viscosity wave observed in the circular chamber. Meanwhile, the observed effect of distinctive solute flow provoked an experimental investigation on its origin.

In the first experiment, the beads were located in the center of cavity B and arranged along the chamber axis (Fig. 6a, inset). Therefore, one bead (black) was closer to the exit of the channel while the other (red) was closer to the rear wall of the cavity. One crystal of sugar was added to the cavity A. The first reaction to the viscosity wave visible as a significant drop in angular velocity was noted after about 40 seconds (Fig. 6a). Surprisingly, the earlier reaction came from red bead. In other words, the bead situated further away from the exit of the channel reacted first. According to the dynamics of standard, uniform diffusion, the sugar should spread in the cavity A, go through the channel and then fill the cavity B starting from the exit of the channel. The experiment was repeated several times and the equivalent, nonintuitive observations were made. The relative size of two beads (one slightly bigger than the other or equally-sized) had no influence on the results. In order to exclude the occurrence of any direction-related rotational effect, the experiment was repeated for the two beads rotating in opposite directions. Reversing the handedness of the optical vortex is done by changing its topological charge to the opposite value, e.g. from +30 to -30. However, the results (Fig. 6b) confirmed previous observations – again, the red bead reacted earlier.

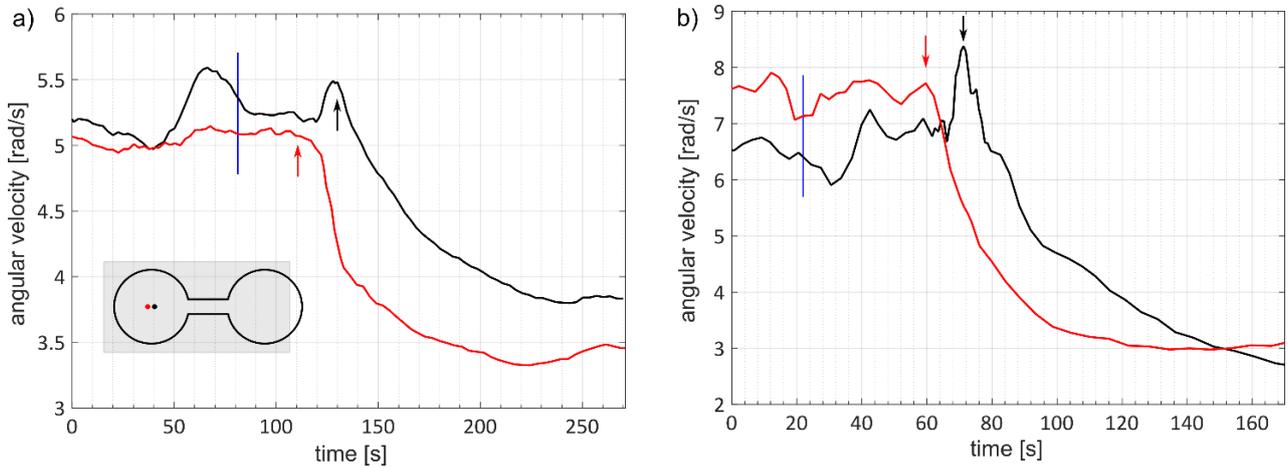

**Fig. 6.** Rotational speed of two beads located in the center of cavity B of bicircular chamber. The approximate positions of the beads (not to scale) are shown in the graphical legend. The moment of one sucrose crystal addition is marked with a blue line and the moments of beads reactions to the viscosity wave were marked with arrows. In both attempts (a) and (b) the drop in rotational speed due to local viscosity increase occurs first for the red bead and then after about 10 s for the black bead. Thus, the red bead (closer to the cavity rear wall) reacted first. The beads rotated in the same direction in experiment (a) and in opposite directions in (b). The increase in velocity at time t = 30 s for black bead in (b) is an artifact – an unwanted shift in depth.

In the second experiment, the beads were located inside the channel and arranged in line perpendicular to the chamber axis (Fig. 7a, inset). This way one bead was closer to the center of the channel while the other was closer to the side wall of the channel. As shown in Fig. 7a, first reaction to the viscosity wave came from the bead situated at the center of the channel (black bead). This observation corresponds to the expected pattern of the laminar flow through the channel.

In the third experiment, the beads were located near the side wall of cavity B and arranged in line perpendicular to the chamber axis (Fig. 7b, inset). It means that one bead was closer to the center of the cavity while the other was closer to its side wall. The outcome of the experiment is presented in Fig. 7b. Soon after the addition of sucrose there is a clear increase in angular velocity due to the mechanical wave followed by the velocity drop at the arrival of the viscosity wave. The earlier reaction was noted for the red, near-the-wall bead, which is again a peculiar observation likewise in the experiment in Fig. 6.

All the bicircular chamber measurements described in the manuscript were rerun 15 times. The reported observations were confirmed in 11 trials. The remaining attempts were inconclusive as low angular velocity of one or both beads made it impossible to judge which bead reacted first. In fact, angular velocity can be regarded as sampling frequency and thus the higher the velocity, the better the time resolution in the measurement.

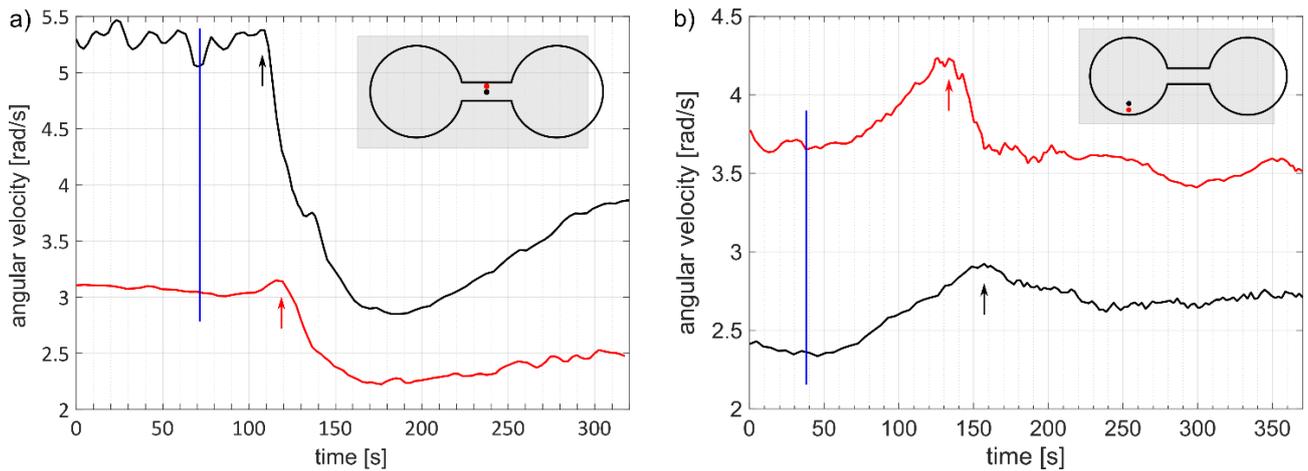

**Fig. 7.** Rotational speed of two beads located (a) in the channel and (b) at the side of cavity B of bicircular chamber. The approximate positions of the beads (not to scale) are shown in the graphical legends. The moment of one sucrose crystal addition is marked with a blue line and the moments of beads reactions to the viscosity wave were marked with arrows. In (a) the drop in rotational speed due to local viscosity increase occurred in t=110 s for black bead and then in t=120s for red bead. Thus, the black bead (farther from the channel wall) reacted first. In (b) drop occurred first for the red bead and then after about 30 s for the black bead. Thus, the red bead (closer to the cavity side wall) reacted first.

## 6. Discussion

It has been speculated that the surprising diffusion dynamics in the bicircular chamber is a heat-related effect stemming from the halogen lamp used as an external white light illumination (see Fig. 2). Even though the measurements were performed in very low illumination intensity, halogen lamp could locally heat up the sample thus inducing convection currents. In order to verify this possibility, halogen lamp was replaced with a cold light source (LED illumination) and the bicircular chamber experiment with both beads in the center of cavity B was repeated (Fig. 8). The arrangement of beads was equivalent to the one in Fig. 6 and the only difference lied in the illumination source. The results remained unchanged – the red bead was first to react to the viscosity wave. Therefore, the thermal origin of the phenomenon was disqualified. Interestingly, the lack of heating visibly slowed down the dissolution process. The beads reacted to sucrose after about 60 seconds while for halogen illumination the reaction time ranged from 30 to 40 seconds.

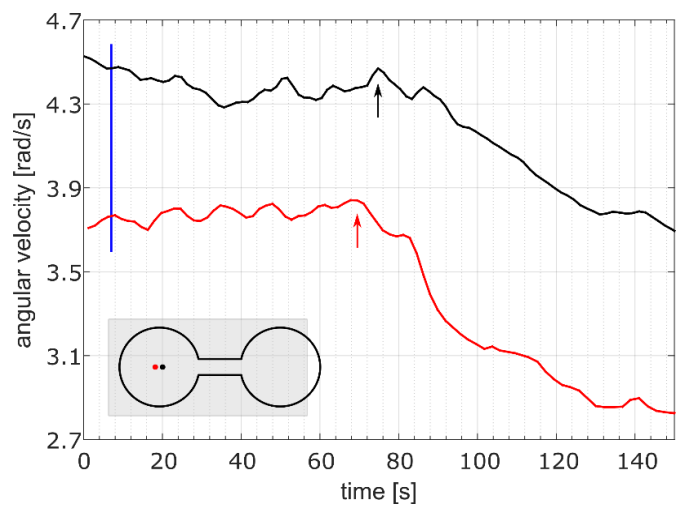

**Fig. 8.** Rotational speed of two beads located in the center of cavity B of bicircular chamber. The approximate positions of the beads (not to scale) are shown in the graphical legend. The moment of one sucrose crystal addition is marked with a blue line and the moments of beads reactions to the viscosity wave were marked with arrows. The experimental configuration differs from the one in Fig. 6 only with the type of illumination. Here, cold light source was used to illuminate the sample. Due to elimination of the heating effect the dissolution of sucrose is slowed down which manifests itself in late response of both beads – first response to sucrose was noted after about 60 seconds compared to 30-40 seconds for "hot" halogen illumination (Fig.6).

It should be noted that the observed effect of nonintuitive diffusion is not limited to sucrose. Analogous experiments were conducted for glycerol – instead of adding one crystal of sucrose, 0.2 µl of 50% aqueous glycerol solution was injected into the cavity A of the bicircular chamber (Fig. 9). The two rotating beads were located at the center of the cavity B and their reactions to were compatible with the ones obtained for sucrose.

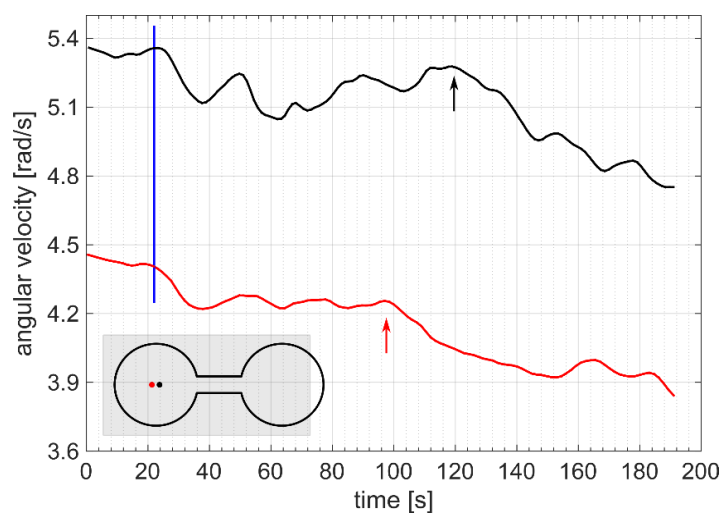

**Fig. 9.** Rotational speed of two beads located in the center of cavity B of bicircular chamber. The approximate positions of the beads (not to scale) are shown in the graphical legend. The moment of injecting 0.2 µl of 50% glycerol solution to the cavity A is marked with a blue line and the moments of beads reactions to the viscosity wave were marked with arrows.

The attempt was made to observe the trajectory of the diffusion directly using two imaging techniques. The first one was phase contrast microscopy [22, 23] suitable for specimens which modulate phase of light instead of its intensity. In principle, the higher sucrose concentration, the higher the refractive index and therefore the stronger phase modulation. Thus, the regions of high and low sucrose concentration should appear differently (in different colors) under the phase contrast microscope. The observations were made at 10x and 20x magnification. Unfortunately. the resolution of the detectable phase change was too low and the diffusion flow pattern could not be determined. Another tested imaging technique was polarized light microscopy [23] dedicated for optically active materials, i.e. materials which rotate the plane of polarization of transmitted light. Optical activity results from asymmetry in material structure (in sucrose molecule due to asymmetric carbon atoms). The angle of the polarization rotation is proportional to sugar concentration and chamber depth. Since the bicircular chamber was only 1 mm deep and one sucrose crystal raised the total concentration by about 2%, the resolution of polarized light microscope was insufficient to observe the ongoing diffusion process.

Although the direct imaging of phase change failed, an alternative approach gave important insight into the diffusion dynamics. During dissolution of sucrose small particles of contaminants, most likely dust from the crystal surface, were noticed under the microscope in the optical tweezers setup. These particles were barely visible at the bottom of the chamber where the metallic beads were located. However, they came into focus when the observation plane lied slightly below the middle depth of the chamber. Refocusing from the bottom to middle surface could not be made with 100x objective used for optical trapping due to its short working distance. Applying 20x objective revealed the stream of particles spreading well above the beads level. Based on this finding, a plausible explanation of the peculiar impression of the reverse solute flow was concluded and sketched in Fig. 10. The stream of dissolved sucrose propagates about 250-300 µm above the bottom glass along the entire length of the chamber. During its way from cavity A to B it causes the forward movement of the nearby layers of water and therefore creates the mechanical wave. This is consistent with the finding that the mechanical wave is driven by the diffusion (see section 5.1). Free beads are carried by the mechanical wave, which is directly observed in the experiment, while trapped beads are displaced from the trap centers into the bright ring, hence their slight angular acceleration. At the same time, the mid-depth solute stream reaches cavity B, gets reflected from the rear chamber wall and splits into upper and lower streams. The lower stream is the above-mentioned viscosity wave which causes sudden drop in beads rotational speed. For beads located in the center of cavity B, the lower stream arrives at their position "from behind", i.e. in the opposite direction to the mid-depth flow, hence the misleading observation of the reverse solute flow. The similar happens for the beads located at the side wall of cavity B. In the case of beads placed in the channel the trend is reversed – the bead further away from the channel wall reacts earlier – due to the slower stream flow near the channel wall (according to the laminar flow).

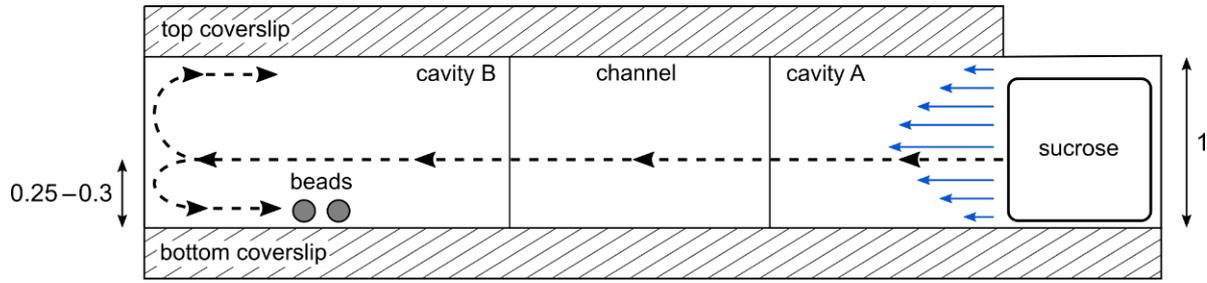

**Fig. 10.** Reproduced diffusion flow in bicircular chamber (not to scale). First, dissolved sucrose spreads at the mid-depth of the chamber from cavity A to cavity B and sets into motion the nearby layers of water (blue arrows), inducing the mechanical wave. The mechanical wave hits the trapped beads "from the front" and speeds up their rotations. Simultaneously, the main solute stream gets reflected from the wall of cavity B and splits into lower and upper streams. The lower stream represents the viscosity wave and it reaches centrally located the beads "from behind", which gives the impression of the reverse solute flow. Units: millimeters.

The temporal resolution of our measurements is somehow dictated by angular velocity of the bead. While measuring the period of each subsequent rotation, the "sampling frequency" and thus the temporal resolution equals the period. This approach is particularly useful while processing the recording by hand. On the other hand, reading the angular bead position more frequently, e.g. every quarter turn or even at each frame, leads to the increase in temporal resolution, but requires the automation of data processing. Another idea to improve the temporal resolution is to induce faster bead rotations, either by increasing the laser power or by using smaller beads (smaller drag force, see Eq. 1). The better the resolution, the more rapid viscosity changes can be detected.

The spatial resolution of our method is limited by the interaction between two beads located close to each other. It has been estimated that the separation of two bead diameters between bead centers is enough to neglect their mutual influence. A similar observation was reported by [24] for beads oscillating along a line (instead of rotating). However, in order to guarantee no interaction, the "safe" distance was assessed at three times the diameter, i.e. about 24-30 μm for 8-10 μm beads. The spatial resolution could be improved by using smaller beads.

## 7. Conclusions

In this paper, two-laser optical tweezers were used for trapping metallic microbeads in optical vortex traps. The trapped beads underwent regular rotations. Their angular velocity (either absolute or relative between two simultaneously rotating beads) became the source of information about the medium surrounding the beads. The method of tracing the direction and speed of solute diffusion process was proposed. The experimental verification of the method concerned the dynamics of sucrose crystal dissolution in two types of measurement chambers. The circular chamber experiments showed the presence of mechanical wave due to the first phase of dissolution process. This wave is followed by the stream of higher sugar concentration, which results in a rapid drop in MB rotational speed. The results obtained for bicircular chamber were surprising and nonintuitive as they indicated the effect of sucrose diffusion in the form of directional stream in the middle depth of the measurement chamber rather than the uniform solute dispersion. The observed effect was related neither to heat transfer nor to the interaction between two MBs. It also applied to glycerol. Unfortunately, the experiment could not be rerun for salts (attempts were made for NaCl and KI) since the chemical interaction between the salt and thin metal coating on the bead leads to the fast degradation of the layer and precludes the measurements.

The provided explanation of the observed effect cannot be considered definitive. Nevertheless, all the repetitive experiments with trapped MBs as well as the movement of free MBs and contaminant particles observed directly under the microscope led us to these conclusions as the most relevant so far.

The bottom line is that the two trapped MBs separated by tens of micrometers can be used to detect minor viscosity changes occurring dynamically in a very small volume of liquid. It may lead to the discovery of some counterintuitive effects as the one reported in this paper.

**Funding** This study was funded by Polish Ministry of Science and Higher Education (grant number DI2016006146).

**Conflicts of interest** The authors declare that they have no conflict of interest.

**Authors' contributions** Conceptualization: Weronika Lamperska and Jan Masajada; Methodology: Weronika Lamperska; Software: Sławomir Drobczyński; Formal analysis and investigation: Weronika Lamperska and Jan Masajada; Writing – original draft, review and editing: Weronika Lamperska, Jan Masajada and Sławomir Drobczyński; Funding acquisition: Weronika Lamperska

**References**

1. Vasnetsov M, Staliunas K (1999) Optical vortices. Nova Science, Huntington

2. Soskin M, Vasnetsov MV (2001) Singular optics. Progress in Optics 42:219-276. https://books.google.pl/books?id=QHaK23Mi8ywC

3. Torres JP, Torner L (2011) Twisted photons: applications of light with orbital angular momentum. Wiley-VCH, Weinheim

4. Padgett M, Bowman R (2011) Tweezers with a twist. Nature Photonics 5:343-348. https://doi.org/10.1038/nphoton.2011.81

5. Jones PH, Maragó M, Onofrio, Volpe G (2015) Optical Tweezers. Principles and applications, Cambridge University Press, Cambridge, England. https://doi.org/10.1017/CBO9781107279711.002

6. Simpson SH, Hanna S (2010) Orbital motion of optically trapped particles in Laguerre–Gaussian beams. JOSA A 27:2061-2071. https://doi.org/10.1364/JOSAA.27.002061

7. Bacia M, Lamperska W, Masajada J, Drobczyński S, Marc M (2015) Polygonal micro-whirlpools in ferrofluids. Optica Applicata 45(3):309-316. https://doi.org/10.5277/oa150304

8. He H, Friese MEJ, Heckenberg NR, Rubinsztein-Dunlop H (1995) Direct observation of transfer of angular momentum to absorptive particles from a laser beam with a phase singularity. Physical Review Letters 75(5):826. https://doi.org/10.1103/PhysRevLett.75.826

9. Parkin S, Knöner G, Nieminen TA, Heckenberg NR, Rubinsztein-Dunlop H (2006) Measurement of the total optical angular momentum transfer in optical tweezers. Optics Express 14:6963–6970. https://doi.org/10.1364/OE.14.006963

10. Asavei T, Nieminen TA, Loke VLY, Stilgoe AB, Bowman R, Preece D, Padgett MJ, Heckenberg NR and Rubinsztein-Dunlop H (2013) Optically trapped and driven paddle-wheel. New Journal of Physics 15(6):063016. https://doi.org/10.1088/1367-2630/15/6/063016

11. Lamperska W, Masajada J, Drobczyński S, Wasylczyk P (2020) Optical vortex torque measured with optically trapped microbarbells. Applied Optics 59(15):4703-4707. https://doi.org/10.1364/AO.385167

12. Lamperska W, Masajada J, Drobczyński S, Gusin P (2017). Two-laser optical tweezers with a blinking beam. Optics and Lasers in Engineering 94:82-89. https://doi.org/10.1016/j.optlaseng.2017.03.006

13. Lindken R, Rossi M, Große S, Westerweel J (2009) Micro-particle image velocimetry (μPIV): recent developments, applications, and guidelines. Lab on a Chip 9(17):2551-2567. https://doi.org/10.1039/B906558J

14. Wereley ST, Meinhart CD (2010) Recent advances in micro-particle image velocimetry. Annual Review of Fluid Mechanics 42:557-576. https://doi.org/10.1146/annurev-fluid-121108-145427

15. Ahangar SB, Konduru V, Allen JS, Miljkovic N, Lee SH, Choi CK (2020) Development of automated angle-scanning, high-speed surface plasmon resonance imaging and SPRi visualization for the study of dropwise condensation. Experiments in Fluids 61(1):12. https://doi.org/10.1007/s00348-019-2844-9

16. Jeong CH, Lee HJ, Kim DY, Ahangar SB, Choi CK, Lee SH (2021) Quantitative analysis of contact line behaviors of evaporating binary mixture droplets using surface plasmon resonance imaging. International Journal of Heat and Mass Transfer 165:120690. https://doi.org/10.1016/j.ijheatmasstransfer.2020.120690

17. Czarske J, Büttner L (2015) Micro Laser Doppler Velocimetry (μ-LDV). In: Li D. (eds) Encyclopedia of Microfluidics and Nanofluidics. Springer, New York, NY. https://doi.org/10.1007/978-1-4614-5491-5_977

18. Langford S, Wiggins C, Tenpenny D, Ruggles A (2016) Positron emission particle tracking (PEPT) for fluid flow measurements. Nuclear Engineering and Design 302:81-89. https://doi.org/10.1016/j.nucengdes.2016.01.017

19. Lee SC, Kim K, Kim J, Lee S, Yi JH, Kim SW, Ha KS, Cheong C (2001) One micrometer resolution NMR microscopy. Journal of Magnetic Resonance 150(2):207-213. https://doi.org/10.1006/jmre.2001.2319

20. Elkins CJ, Alley MT (2007) Magnetic resonance velocimetry: applications of magnetic resonance imaging in the measurement of fluid motion. Experiments in Fluids 43(6):823-858. https://doi.org/10.1007/s00348-007-0383-2

21. Landau LD, Lifshitz EM (1987) Fluid Mechanics. Volume 6 of Course of Theoretical Physics, 2nd ed., Pergamon Press, Oxford

22. Burch CR., Stock JPP (1942) Phase-contrast microscopy. Journal of Scientific Instruments 19(5):71. https://doi.org/10.1088%2F0950-7671%2F19%2F5%2F302

23. Pluta M (1988) Advanced Light Microscopy. PWN, Warsaw

24. Curran A, Lee MP, Padgett MJ, Cooper JM, Di Leonardo R (2012) Partial synchronization of stochastic oscillators through hydrodynamic coupling. Physical Review Letters 108(24):240601. https://doi.org/10.1103/PhysRevLett.108.240601

SUPPLEMENTARY INFORMATION

for

**Microscale solute flow probed with rotating microbead trapped in optical vortex**

Weronika Lamperska[1,*], Jan Masajada[1], Sławomir Drobczyński[1]

[1]Department of Optics and Photonics, Faculty of Fundamental Problems of Technology,
Wroclaw University of Science and Technology, Wybrzeże Wyspiańskiego 27, 50-370, Wrocław, Poland;
*Corresponding author: weronika.lamperska@pwr.edu.pl

**Details on viscosity calculations and the influence of evaporation**

Let us consider a circular chamber filled with 78 μl of water. The average mass of a single sucrose crystal was estimated at (1.0±0.1) mg. Before the addition of the first sucrose crystal the concentration is 0% (pure water). By adding subsequent sucrose crystals the concentration (w/w) should be equal to: (1.27±0.12)% for 1 crystal, (2.50±0.24)% for 2 crystals and (3.70±0.35)% for 3 crystals. These concentrations refer to the solutions of fully dissolved sucrose. Viscosity of aqueous sucrose solutions at different concentrations at 25°C can be found in literature [1]. In order to obtain the viscosity values for particular concentrations, the literature data points were fitted with a biexponential function (Fig. SI.1) of a form

$$f(x) = a \exp(bx) + c \exp(dx) \tag{SI.1}$$

The fitted coefficients are: $a = 0.001788$; $b = 0.1651$; $c = 0.7892$; $d = 0.03901$. Given the equation, one can calculate the viscosity of 1, 2 and 3-crystal solution. For this paper it is more convenient to use relative viscosity, i.e. viscosity of a given solution divided by the viscosity of water.

**Table SI.1.** Literature data for the viscosity of sucrose solutions at different concentrations (w/w) at 25°C. Source: [1] except for *[2].

| sucrose concentration [%, w/w] | viscosity [mPa·s] |
|---|---|
| *0 | 0.89 |
| 20 | 1.695 |
| 30 | 2.735 |
| 40 | 5.164 |
| 50 | 12.40 |
| 60 | 43.03 |

The water evaporates during the experiment which may have noticeable impact on the measurements in case of long-lasting experiments, such as one-bead experiment in circular chamber (almost 23-minute long). As mentioned in the manuscript (Section 5.1) evaporation starts at the gap in the chamber (Fig. SI.2a) and progresses in a vertical direction until a substrate becomes dry. Further evaporation gives rise to the concave meniscus which propagates horizontally in the fluid layer (Fig. SI.2b). The meniscus can be observed with the naked eye (Fig. SI.2c). It takes about 20 minutes until the meniscus starts to form. It means that the volume of water in the open area evaporates in 20 minutes. For a 1.5 mm wide gap the volume of the open region is about 9.4% of the total chamber volume. Thus, the evaporated volume is about 7.3 μl. Assuming that evaporation rate is constant in time, one can calculate the water loss at any moment of the experiment. Let us consider the first sucrose addition. In Fig. 5a first crystal was added in the 188. second of the experiment. The viscosity undergoes a sudden jump and then starts to lower and stabilize. Just before the addition of the second crystal (492. second) the viscosity should be closest to the literature value (blue strips in Fig. 5b) since the dissolution has almost or completely finished. However, until then about 2.9 μl of water have already evaporated. One needs to recalculate literature values for the slightly increased concentration (dashed red line in Fig. 5b). The more sucrose added and the longer the experiment, the larger discrepancy between "with" and "without evaporation" case. It is worth mentioning that the described evaporation rate holds for halogen lamp illumination producing significant amounts of heat ("hot" light source). Further in the manuscript halogen lamp was replaced with a LED source ("cold" light source). Then, the evaporation slows down more than twice – the meniscus starts to arise after about 45 minutes.

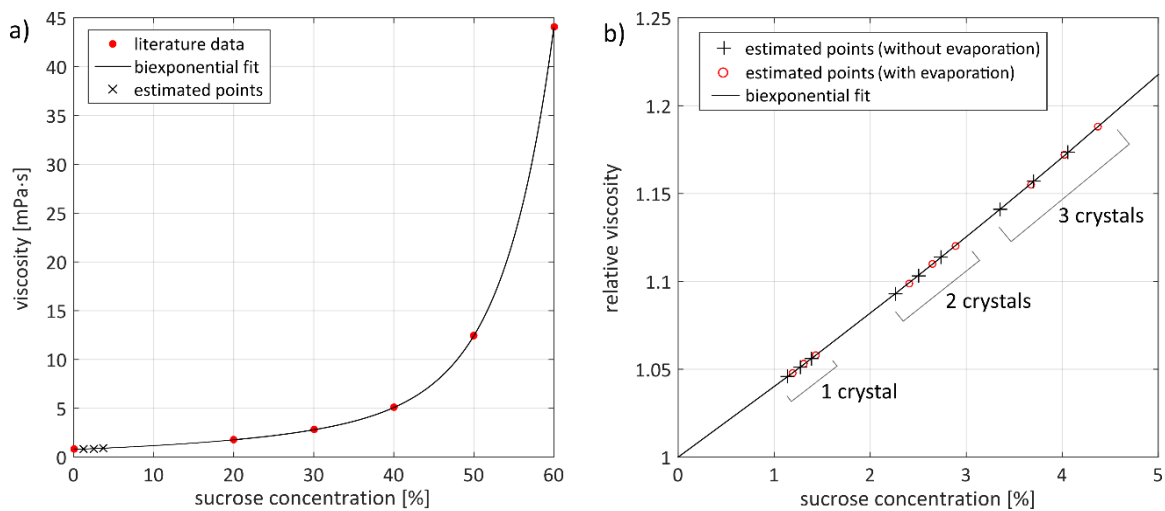

**Fig. SI.1.** Biexponential function (solid black line) $f(x) = 0.001788\exp(0.1651x) + 0.7892\exp(0.03901x)$ fitted to the literature data (from Table SI.1; red dots); (a) The evaluated sucrose concentrations for average mass of 1-3 sucrose crystals dissolved in a circular chamber were marked with 'x' signs; (b) Relative viscosity for small sucrose concentrations based on the fit. The viscosity for estimated sucrose concentrations with neglected evaporation were marked with '+' signs. There are three points for each crystal addition representing three distinctive cases: the lowest concentration (each crystal of 0.9 mg), the average concentration (each crystal of 1.0 mg) and the highest concentration (each crystal of 1.1 mg). Analogously, the viscosity for estimated sucrose concentrations with evaporation included were marked with red circles.

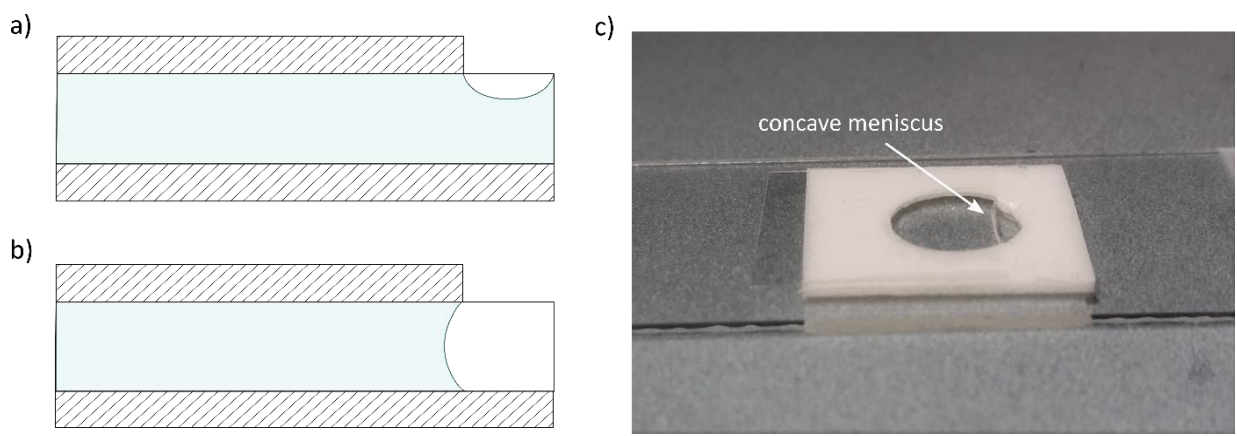

**Fig. SI.2.** Evaporation in the circular chamber; (a) evaporation starts at the gap region and continues downwards the chamber; (b) after the entire open area gets dry, the concave meniscus appears and propagates horizontally into the fluid layer; (c) the photo of a meniscus in the sample. Here, instead of two coverslips there is one coverslip (top) and one standard microscope slide (bottom).

**References**

1. Swindells JF, United States National Bureau of Standards (1958) Viscosities of Sucrose Solutions At Various Temperatures: Tables of Recalculated Values, Washington

2. Venard JK, Street RL (1975) Elementary Fluid Mechanics, 5th ed., Wiley, New York